\documentclass[conference, letterpaper]{IEEEtran}
\IEEEoverridecommandlockouts
\usepackage{cuted}
\usepackage{cite}
\usepackage{bm}
\usepackage{amsmath,amssymb,amsfonts}
\usepackage{algorithm}
\usepackage{algorithmic}
\usepackage{graphicx}

\usepackage{graphicx}
\usepackage{subcaption}
\usepackage{makecell}
\usepackage{diagbox}

\usepackage[letterpaper, left=0.625in, right=0.625in, bottom=1.02in, top=0.70in]{geometry}

\usepackage{textcomp}
\usepackage{xcolor}

\newcommand{\be}{\begin{equation}}
\newcommand{\ee}{\end{equation}}

\def\BibTeX{{\rm B\kern-.05em{\sc i\kern-.025em b}\kern-.08em
    T\kern-.1667em\lower.7ex\hbox{E}\kern-.125emX}}

\begin{document}

\title{Map2Schedule: An End-to-End Link Scheduling Method for Urban V2V Communications
}

\author{Lihao Zhang,  Haijian Sun, Jin Sun, Ramviyas Parasuraman,  Yinghui Ye, Rose Qingyang Hu

\thanks{
L. Zhang and H. Sun are with the School of Electrical and Computer Engineering, University of Georgia, Athens, GA, USA.}
\thanks{J. Sun and R. Parasuraman are with the School of Computing, University of Georgia, Athens, GA, USA.}
\thanks{Y. Ye is with Shanxi Key Laboratory of Information Communication Network and Security, Xi’an University of Posts \& Telecommunications, China.}
\thanks{R. Q. Hu is with the Electrical and Computer Engineering Department at Utah State University, Logan, UT, USA.}
}

\maketitle

\begin{abstract}
\par Urban vehicle-to-vehicle (V2V) link scheduling with shared spectrum is a challenging problem. Its main goal is to find the scheduling policy that can maximize system performance (usually the sum capacity of each link or their energy efficiency).  Given that each link can experience interference from all other active links, the scheduling becomes a combinatorial integer programming problem and generally does not scale well with the number of V2V pairs. Moreover, link scheduling requires accurate channel state information (CSI), which is very difficult to estimate with good accuracy under high vehicle mobility. 
In this paper, we propose an end-to-end urban V2V link scheduling method called Map2Schedule, which can directly generate V2V scheduling policy from the city map and vehicle locations. Map2Schedule delivers comparable performance to the physical-model-based methods in urban settings while maintaining low computation complexity. This enhanced performance is achieved by machine learning (ML) technologies. Specifically, we first deploy the convolutional neural network (CNN) model to estimate the CSI from street layout and vehicle locations and then apply the graph embedding model for optimal scheduling policy. The results show that the proposed method can achieve high accuracy with much lower overhead and latency.  
\end{abstract}


\section{Introduction}
\par As vehicles become increasingly intelligent, thanks to powerful onboard chips and sensors, it is becoming both inevitable and progressively feasible to enable them to establish connections with other vehicles (V2V), infrastructure (V2I), and pedestrians (V2P) for advanced transportation applications.  This is the core concept of vehicle-to-everything (V2X) technologies\cite{7992934}, which aims to enhance the transportation system to become more comfortable, environment-friendly, efficient, and safer than ever. Among the various types of V2X communications, V2V plays a crucial role due to its potential to facilitate autonomous driving and provide timely incident alerts. Nevertheless, a significant portion of V2V communications will occur in dense urban environments, potentially resulting in substantial mutual interference and calling for appropriate link scheduling mechanisms.
\par To better utilize the wireless resources, a common practice is to divide the wireless channels into small resource blocks in either the time or frequency domain. While this can help mitigate interference, spectrum utilization may not be optimal and can potentially be further enhanced.  Dynamic Spectrum sharing, on the other hand, allows multiple users to access the same spectrum resources. The active users are allowed to share the spectrum resources when their mutual interference is deemed low. As a result, many optimal scheduling algorithms have been developed using diverse mathematical frameworks.  Nonetheless, these algorithms tend to suffer from the drawback of high computational complexity. For example, \cite{qian2010s} proposed a global-optimal scheduling algorithm called \textit{S-MAPEL}, which has a high complexity with respect to pairing the links.
\par To mitigate the concern of high computational complexity, certain sub-optimal algorithms have been developed. One of the early contributions in this area is the greedy heuristic search algorithm \textit{FlashLinQ}\cite{wu2013flashlinq}. Authors in \cite{geng2015optimality} proposed a general framework that link scheduling is optimal for the whole generalized-dimension-of-freedom (GDoF) region and within a constant gap of the capacity region when it is subject to the treat interference-as-noise  (TIN) condition. Based on TIN, \textit{ITLinQ}\cite{naderializadeh2014itlinq} and \textit{ITLinQ+} \cite{yi2016optimality} were then developed to find the sub-optimal scheduling policy by sequential link selection. However, the performance results of \textit{FlashLinQ}, \textit{ITLinQ}, and \textit{ITLinQ+} heavily depend on empirical parameter design. In \cite{shen2017fplinq}, the authors proposed another iterative algorithm, \textit{FPLinQ}, derived from fractional programming. Although this algorithm outperforms the above algorithms with no need to do parameter fine-tuning, its iterative process still reveals high computation complexity that can hardly meet real-time V2V communication requirements under high mobility.

\par Furthermore, it is worth noting that the above methods are based on the assumption that full CSI (including direct and interference channels) is available. The direct CSI for a transmitter-receiver pair can be acquired during their communication process. However, the link scheduling problem  not only needs the direct CSI but also needs the information about all the interfering channels, a task that proves to be quite challenging and time-consuming in the high-mobility scenarios.  With the surge in machine learning (ML) technologies, some end-to-end ML-based scheduling methods have been developed. In \cite{cui2019spatial}, the authors proposed ``spatial deep learning", an end-to-end ML approach that generates link scheduling based on spatial distances. In \cite{lee2020graph}, the authors proposed a fast link scheduling based on the graph embedding model by treating the V2V links as graphs.  However, the above schemes  only considered transmitter-receiver pair distance during scheduling, which may work well in rural, wild, or flat environments as distance has a strong correlation with CSI.  Distance based scheduling can lead to significant scheduling policy deviations in urban V2V scenarios where buildings and foliage are dense.


To achieve real-time and optimal V2V link scheduling, it is essential to have both accurate full CSI estimation and low-complexity scheduling algorithms in place. Furthermore, it is crucial that the scheme needs to scale well with an increasing number of links. 
In this work, we explore a new approach driven by ML and present the following contributions. 
\begin{itemize}

\item We develop an end-to-end link scheduling method named \textit{Map2Schedule}, which can directly generate a scheduling policy that requires only city maps and vehicle locations, thereby significantly reducing signal overhead and computation complexity. 

\item \textit{Map2Schedule} is trained on physical-model-based simulation data and performs $80\%$ higher on sum-rate metrics compared to the existing methods such as \cite{cui2019spatial} and \cite{lee2020graph}. Meanwhile, \textit{Map2Schedule} exhibits low latency and can schedule 50 V2V links within 0.2 seconds, a reduction of two orders of magnitude compared to our performance baseline. This runtime was tested without any optimization applied to the computation process. Besides, the proposed algorithm can handle an arbitrary number of V2V links, which can greatly facilitate transfer learning and practical implementation.


\item We have also explored transfer learning for better adaptation to unforeseen scenarios. The results show that our model has good transferability and can quickly adapt to new tasks with a few-shot method.

\end{itemize}

The rest of the paper is organized as follows. In section \uppercase\expandafter{\romannumeral2},  the end-to-end wireless link scheduling problem is presented and the problem is divided into two sub-problems. Section \uppercase\expandafter{\romannumeral3} introduces the dataset and the system model. In section \uppercase\expandafter{\romannumeral4}, the  experiment settings and the results are presented. Finally, this paper is concluded in section \uppercase\expandafter{\romannumeral5}. 

\begin{figure}[h]
    \centering
    \includegraphics[width=0.55\linewidth]{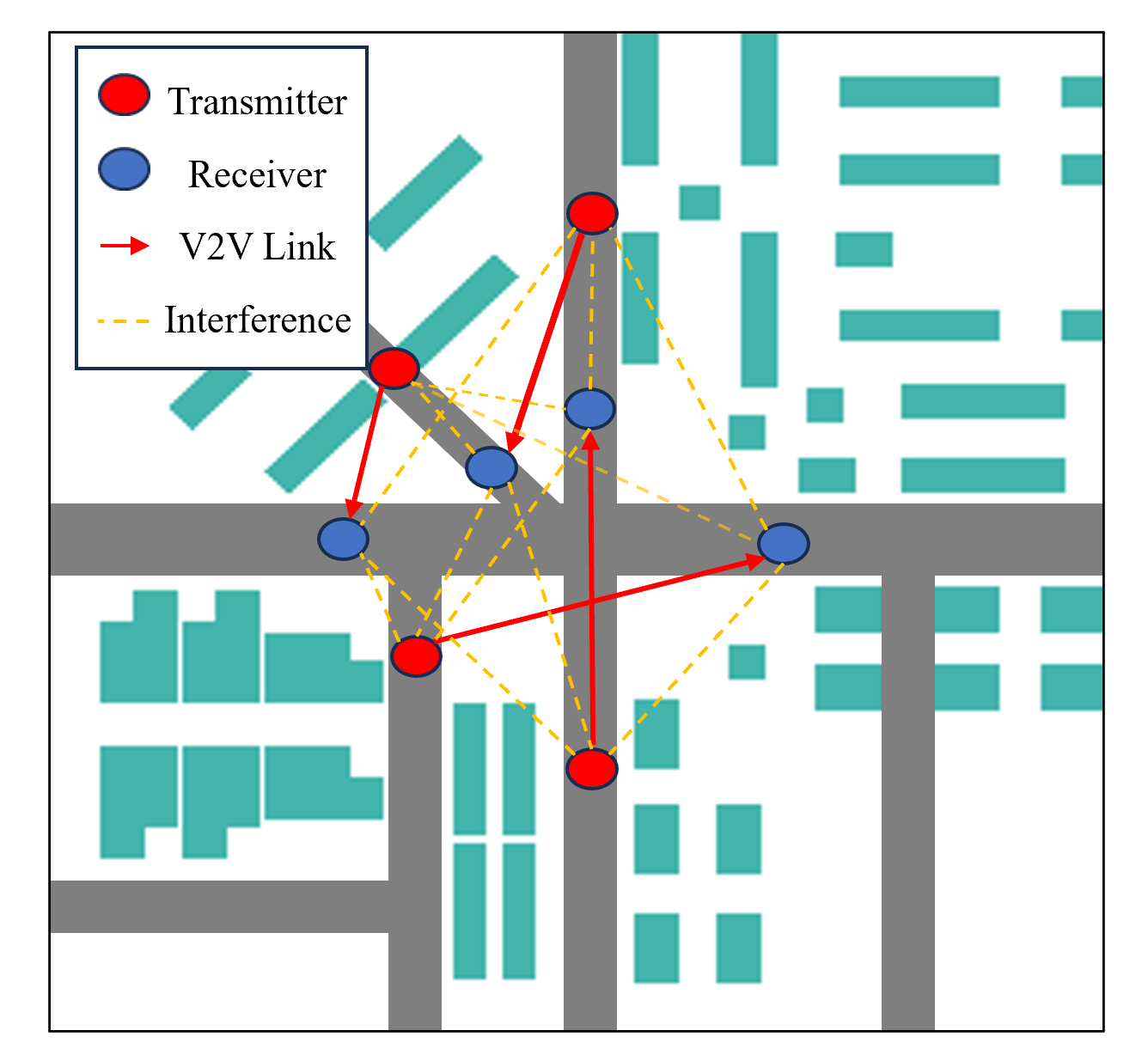}
    \caption{A typical urban V2V scenario.}
    \label{fig:v2v}
\end{figure}

\section{Background and Problem Formulation}

As shown in the Fig. \ref{fig:v2v}, we consider a typical urban V2V scenario which consists of a city map and $N$ overlaid wireless links denoted as $d_{i}\in\mathcal{D}\{d_{1},d_{2},d_{3},\cdots\ , d_N \}$.  The gray areas are roads and the green ones are buildings. Each $d_{i}$ refers to a pair of a transmitter $t_{i}$ and its corresponding receiver $r_{i}$. Consequently, link set $\mathcal{D}$ corresponds to $N$ transmitters $t_{i} \in \mathcal{T}\{t_{1},t_{2},t_{3}\cdots\ t_{N}\}$ and $N$ receivers $r_{i} \in \mathcal{R}\{r_{1},r_{2},r_{3}\cdots\ r_{N}\}$. To boost spectrum efficiency, all the links share the whole frequency resource. We denote $h_{ii}$ as the CSI from $t_{i}$ to $r_{i}$, also is the direct channel gain of link $d_{i}$, and denote $h_{ji}$ for CSI from $t_{j}$ to $r_{i}$, also is the interference gain from link $d_{j}$ to link $d_{i}$.

\par In the considered urban V2X scenario, we aim to find the optimal link scheduling policy (on or off state of each $d_{i}$) to maximize the system sum-rate. In this paper, we only schedule the links to either be activated with power $p_{i}$ or be turned off, which is represented by a binary indicator $x_{i}\in\{0,1\}$ for link $d_{i}$. Without loss of generality, we assume all links have the same power $p_{i}$. $\mathbf{x}=[x_1,x_2,\cdots,x_N]$ denotes the link scheduling policy of $\mathcal{D}$.   The signal-to-interference plus noise ratio (SINR) of $d_{i}$ can be written as Eq. \eqref{sinri} where the $\sigma^2_N$ denotes the additive white Gaussian noise (AWGN), and the communication rate of $d_{i}$ can be written as Eq. \eqref{capacity} where $B$ is the system bandwidth.
\begin{equation}
SI\!N\!R_i(\mathbf{x}) = \frac{x_ip_i\lvert h_{ii}\rvert^2} {\sigma^2_N + \sum_{j\neq i}x_jp_j\lvert h_{ji}\rvert^2},
\label{sinri}
\end{equation}
\begin{equation}
R_i(\mathbf{x}) = B \log_2(1 + SINR_i(\mathbf{x})).
\label{capacity}
\end{equation}
From Eq. \eqref{capacity}, we could find that the communication rate of $d_{i}$ is impacted by interference. The optimization problem can be formulated as \eqref{problem}, which means we should find a proper scheduling policy $\mathbf{x}$ that only activates the ``good" links with high channel gain and low interference to others.
\begin{equation}
\begin{aligned} \label{problem}
&\mathop{max}\limits_{\mathbf{x}}\sum\limits_{i=1}^NR_i(\mathbf{x}) \\
&s.t.\quad x_i \in \{0, 1\}, \forall x_i \in \mathbf{x}.
\end{aligned}
\end{equation}
This problem naturally splits into two sub-problems. The first sub-problem is to obtain $h_{ij},\forall i,j\in \{1,2\ ...\ N \}$, the $N\times N$ CSI values between all possible transmitter-receiver pairs. The second sub-problem is to generate the link scheduling policy based on the CSI values.

\begin{figure*}[ht]
    \centering
    \includegraphics[width=0.9\linewidth]{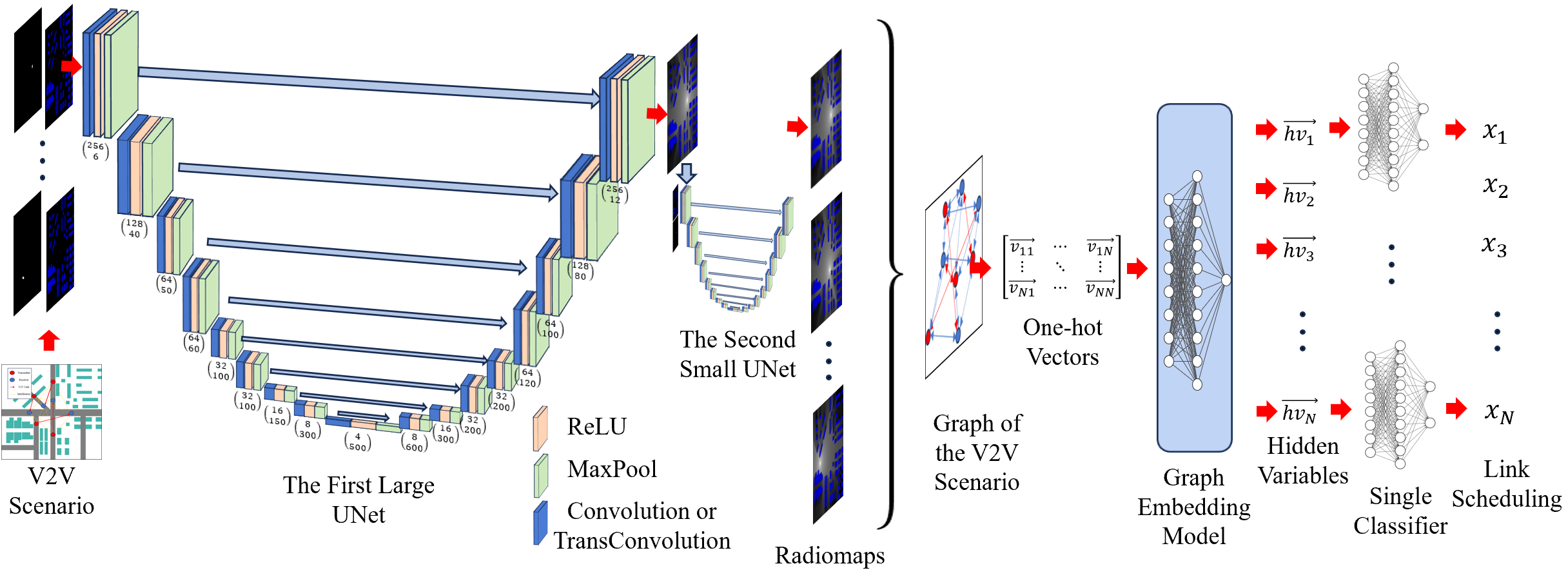}
    \caption{Map2Schedule structure}
    \label{fig:sysmodel}
\end{figure*}

\subsection{Channel Estimation}

In practice, CSI is obtained by periodically sending pilot signals. However, they experience high overhead and latency, not to mention implementing a centralized algorithm for indirect CSI pairs. 
In the literature, one of the most accurate approaches is physical-model-based simulations, like the dominant path model (DPM)\cite{wahl2005dominant} and ray tracing\cite{rizk1997two}. Although such simulations can provide precise CSI prediction, it will take minutes to generate full CSI. In the past few years, some ML-based CSI predicting methods emerged. However, most are developed for single transmitter-receiver pairs, like\cite{saito2019two,6476630,imai2019radio}. 

Such single-pair methods are not applicable to large-scale V2V cases, where the number of links can reach to hundreds. Fortunately, with recent advancements in ML research, new deep learning methods were developed and can emulate the physical model with lower computation complexity.

\subsection{Link Scheduling}
\par The second sub-problem is to find optimal or sub-optimal link scheduling policy based on available CSI. In fact, given all $h_{ij}$,  Eq. (\ref{problem}) is a classical optimization problem. Traditional scheduling algorithms use various optimization techniques to find the optimal $\mathbf{x}$. As discussed in the introduction part, \textit{S-MAPEL} \cite{qian2010s}, \textit{FlashLinQ}, \textit{ITLinQ}, \textit{ITLinQ+}, and \textit{FPLinQ}  represent past decade's efforts in this field. Nevertheless, they all suffer from high computation complexity and poor scalability, hence are not suitable for the considered V2V scenarios. 

Recently, some ML-based scheduling algorithms with low computation complexity have been proposed. The closest works are \cite{cui2019spatial} and \cite{lee2020graph}. Both works are end-to-end scheduling predictions and only take link distance as a system input. Their performance builds on the implicit correlation between distance and CSI and, therefore will not work well where distance and channel have large deviations, such as in urban environments. Besides, they can also suffer from poor generalizability, which hinders practical applications.


\section{Scheduling Via Map2Schedule}
We present the details of the proposed  \textit{Map2Schedule} approach, which is designed to generate scheduling policy $\mathbf{x}$ directly from information provided in Fig. \ref{fig:v2v}. The proposed system is illustrated in Fig. \ref{fig:sysmodel}. The first component is a modified CNN model from \textit{RadioUNet} \cite{levie2021radiounet}, which predicts the CSI from the 2-dimensional (2-D) city map and vehicle locations. The second component is the graph embedding model, which uses the predicted CSI to generate a near-optimal scheduling policy in real time.

\subsection{Dataset}
\begin{figure}[h]
    \centering
    \includegraphics[width=0.8\linewidth]{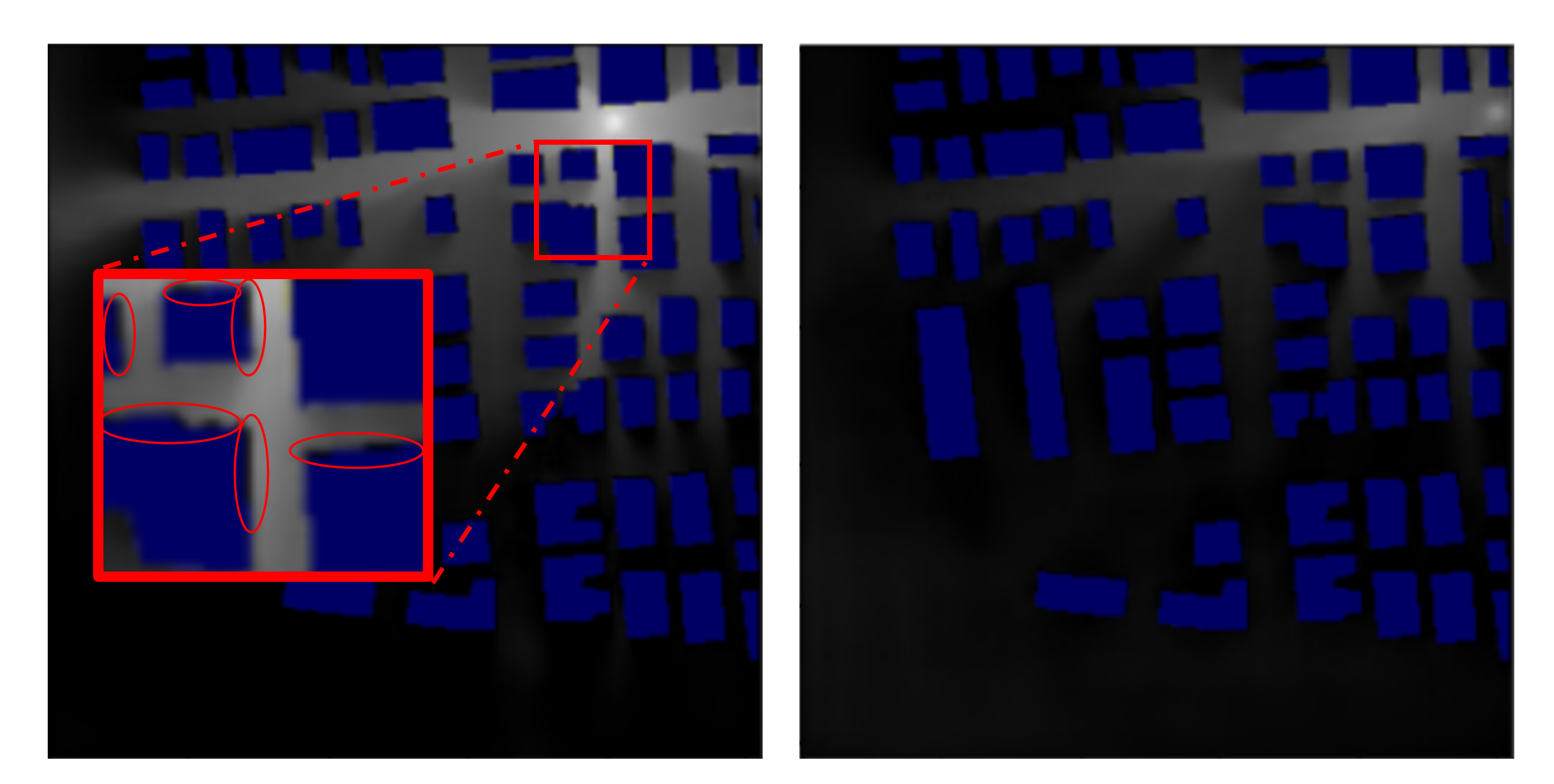}
    \caption{Black edge and transmitter fading}
    \label{fig:radiounet_fault}
\end{figure}

In this paper, our dataset consists \textit{RadioMapSeer} \cite{levie2021radiounet} and the near-optimal scheduling policy $\mathbf{x}$ generated by \textit{FPLinQ}. \textit{RadioMapSeer} is a dataset of 56,000 radiomaps. As shown in Fig. \ref{fig:radiounet_fault}, each radiomap is an image of the signal strength distribution, and a brighter pixel corresponds to a stronger signal at that location. These radiomaps were simulated on 700 2-D city maps. For each map, there are 80 radiomaps simulated by WinProp with the DPM algorithm, corresponding to 80 transmitter locations. The map size is 256 m $\times$ 256 m. 

\par For our scheduling task, we built V2V scenarios from the \textit{RadioMapSeer} dataset. These scenarios were configured based on two key parameters: the number of links and the range of link distances. The number of links is configured as 10, 20, 30, 40, and 50, and the range of link distance is configured as 2 meters to 32 meters for short-distance groups and 2 meters to 65 meters for long-distance ones. We extracted the CSI matrix from the radiomap and then utilized the \textit{FPLinQ} algorithm to obtain link scheduling policy as the baseline (ground truth). 

\vspace{-0.2cm}

\subsection{Radiomap Prediction}
To predict the CSI, we deployed the \textit{RadioUNet}, a CNN model based on the \textit{UNet} architecture. \textit{UNet} architecture is symmetric and with similar shape as letter ``U'' as shown in Fig. \ref{fig:sysmodel}. This architecture consists of multiple levels, with two network blocks on each level. The left-side blocks perform convolution (down-sampling), activation, and pooling, generally used as feature extracting or encoding. On the right side, corresponding right-side blocks will do transposed convolution (up-sampling), activation, and pooling, acting as feature expanding or decoding. Unlike other CNN models, the prominent feature of \textit{UNet} is the encoder-decoder path on each level as the blue long arrows in Fig. \ref{fig:sysmodel}. These paths will send the output of the left-side blocks into the input of the right-side blocks. 

\par Furthermore, \textit{RadioUNet} has two connected \textit{UNet} architectures. Both of the \textit{UNet} architectures have ten levels, but the first \textit{UNet} has more channels than the second one. The inputs of the first \textit{UNet} consist of one 2-D city map and one transmitter location image. Its output is a coarse radiomap prediction. After that, the inputs of the second one \textit{UNet} are nearly identical to the first \textit{UNet}, except the output of the first \textit{UNet} is added. 


Original \textit{RadioUNet} is trained for just one transmitter. We have modified its input structure to support multiple direct pairs. Since all links use the same frequency, $N$-pair CSI can be obtained by stacking up each individual radiomap and fine-tuning the output layer. Besides, original \textit{RadioUNet}  exhibits two issues. The first issue was black edges as illustrated in left side of Fig. \ref{fig:radiounet_fault}, where noticeable black edges surround the ``bright faces" of buildings. These black edges correspond to mispredicted zero channel gain at that location. One conjecture is that the model has learned the pattern of black edges associated with ``dark faces'' of buildings and mistakenly applied this knowledge to ``bright faces'' during prediction. To rectify this issue, a straightforward solution is to detect abrupt changes in non-building areas and replace the abnormal pixels with their largest neighbors.

The second issue is ``transmitter fading'' as shown on the right side of Fig. \ref{fig:radiounet_fault}. The transmitters in edge areas significantly ``fades". This issue can be attributed to biased data in the training set, where there are few transmitters located at the map edges. Methods to mitigate data bias include adding compensatory data and data augmentation, such as segmenting the original radiomaps to generate additional samples with transmitters at the edges. In our implementation, we addressed this issue by dropping transmitters at selected locations.

\subsection{Graph Embedding Link Scheduling}
\par Essentially, the link scheduling is to determine the relationship between the direct CSI ($h_{ii}$) and the interference to all other receiver nodes ($h_{ij}, j \neq i$). Such a model can naturally connect with graph structure. In this paper, we applied and modified a popular graph embedding model \textit{structure2vec} \cite{dai2016discriminative}, which is a powerful method for data with graph structure. Its main idea is constructing embedded hidden variables via a nonlinear learnable function for each graph data point (each node in the graph). These hidden variables represent the whole graph from the perspective of each node, which means these variables encapsulate information about the node itself and its relationship to the whole graph. Depending on the problem objectives, these embedded hidden variables could be processed by classifiers, regression models, or other appropriate models. After determining the whole structure, the embedding learnable function and the subsequent task model will be optimized simultaneously through each backward propagation process.

\par To implement the graph embedding model on link scheduling tasks, the first step is to represent each V2V scenario as a graph. While it might seem intuitive to treat each transmitter and receiver as a node, doing so will result in each node missing its node feature. This is unsuitable for graph embedding models that highly depend on the node feature to embed. Considering that our task is to find ``good" links, treating each link as a node is a more natural idea and will make it easier to design the subsequent classifier.

Denote $G(V,E,\alpha)$ as the directed graph of all the links in one V2V scenario, $v_i\in V$ denotes the node $v_i$ in the nodes set $V$, which represents the link $d_i$. $e_{ij}\in E$ is the edge from node $v_i$ to node $v_j$. $\alpha_{ij}$ denotes the edge feature of $e_{ij}$ which is equal to $h_{ij}$, and $\alpha_{ii}$ denotes the node feature of $v_i$ which is equal to $h_{ii}$. From the graph embedding theory, the embedded hidden variable $\mu_i$ for node $v_i$ will be updated iteratively as Eq. \eqref{original iter}. $N(v_i)$ is the set of neighbors of $v_i$.
\begin{equation}
\mu_i^{(t+1)} = \Gamma(\alpha_{ii},\{ \alpha_{ji}\}_{j|v_j\in N(v_i)},\{ \mu_{v_j}^{(t)}\}_{v_j\in N(v_i)}).
\label{original iter}
\end{equation}

Here, $\Gamma$ represents the nonlinear learnable function for inferring the hidden variables, and there are several approximate inference algorithms. In this paper, we choose the mean-field inference algorithm, as shown in Eq. \eqref{mean field}.
\begin{equation}
\mu_i^{(t+1)} = \sigma(W_1\alpha_{ii}+W_2\sum_{j|v_j\in N(v_i)}\alpha_{ji}+W_3\sum_{j|v_j\in N(v_i)}\mu_j^{(t)}).
\label{mean field}
\end{equation}

\par To facilitate model computation,  the CSI is quantized to $p$-dimensional one-hot code. This quantization process compresses the continuous CSI feature space into discrete $p$ categories. Since we aim to perform binary classification on each link, the $p$ categories of CSI should both provide sufficient accuracy and enable faster link scheduling learning. In this paper, the node and edge features are embedded into 8-dimensional one-hot codes, the hidden variable are iterated  twice, and  the embedded hidden variables are represented as 32-dimensional vectors. Consequently, the shapes of $W_1$ and $W_2$ in Eq. \eqref{mean field} are both $32 \times 8$, while $W_3$ is $32 \times 32$. The above setting of hyperparameters was found to be optimal through experiments in \cite{lee2020graph}.
\par After the graph embedding model is constructed, the classifier for each hidden variable can be built. The classifier contains just one hidden variable layer. So, the size of the first weight  is $32 \times 64$, and the second is $64 \times 2$. It may appear counter-intuitive that the classifier only processes one hidden variable at a time, lacking a global view. However, from Eq. \eqref{mean field}, we can find that each embedded hidden variable is calculated from the whole graph, and the CSI values are formulated similarly to \textit{FPLinQ}, which means those hidden variables should be able to mimic how \textit{FPLinQ} algorithm does link scheduling. Meanwhile, the single classifier, as well as the previous embedding model, can accommodate any number of inputs. Hence, the proposed system architecture can accommodate an arbitrary number of V2V links without requiring any modifications. This capability offers significant advantages for practical implementation, transfer learning, and more.

\subsection{Training Process}

\par The proposed  system is trained as two interconnected components in a supervised fashion. For the first component, the CNN model is trained using the \textit{RadioMapSeer} dataset. As mentioned before, the 56,000 radiomaps were simulated on 700 city maps, with 80 different transmitter locations on each map. We divide the radiomaps based on different city maps, allocating them as follows: 400 for the training set, 100 for the validation set, and 200 for the test set. The mean square error (MSE) loss function is employed and the Adam optimization algorithm is  used with a learning rate of $10^{-4}$. 
\par In the second component, the training input consists of V2V CSI extracted from DPM-simulated radiomaps, and the target link scheduling policy is generated by the \textit{FPLinQ} algorithm, serving as the ground truth.
The dataset splits follows  800/1000/4000 for train/validation/test respectively. The reason for such a small training set is as follows. The work in \cite{lee2020graph} shows shows that training on a small dataset has a similar performance compared with large datasets. 
\par Moreover, our problem involves a smaller map size, the links cluster and more overlap in these scenarios, which lead to low link activation ratio and make our problem more challenging. Due to the low activation ratio, the original graph embedding model can not learn the pattern quickly and precisely, as shown in the result section. For example, the prediction of sum-rate can only achieve about $60\%$ of ground truth \textit{FPLinQ} performance. This inconsistency comes from the biased data of much more ``bad" links in each V2V scenario. Because of the utilization of the normal binary cross entropy (BCE) loss function, the model will focus on all input data samples equivalently and will be trained more by the ``bad" link samples. Consequently, the trained model will have low accuracy on ``good" links, which is more important to the sum-rate performance in our scenarios.
\par To alleviate the bias, we deployed the weighted cross entropy loss function. By adjusting the weight of the classes, the accuracy of good links weighs more in the loss function. It leads the parameters of the model descent along the direction that improves the accuracy on good links and the sum-rate metric. In addition, given the variations in the CSI distribution caused by different link numbers and ranges of link distance, we conducted the grid search for optimal training hyperparameters on each scenario setting.

\section{Experimental Results}
Here, we present the results of our experiments. The training process and test experiments were implemented in PyTorch.

\subsection{Sum-rate Performance}
\par
\begin{figure}[h]
    \centering
    \includegraphics[width=0.9\linewidth]{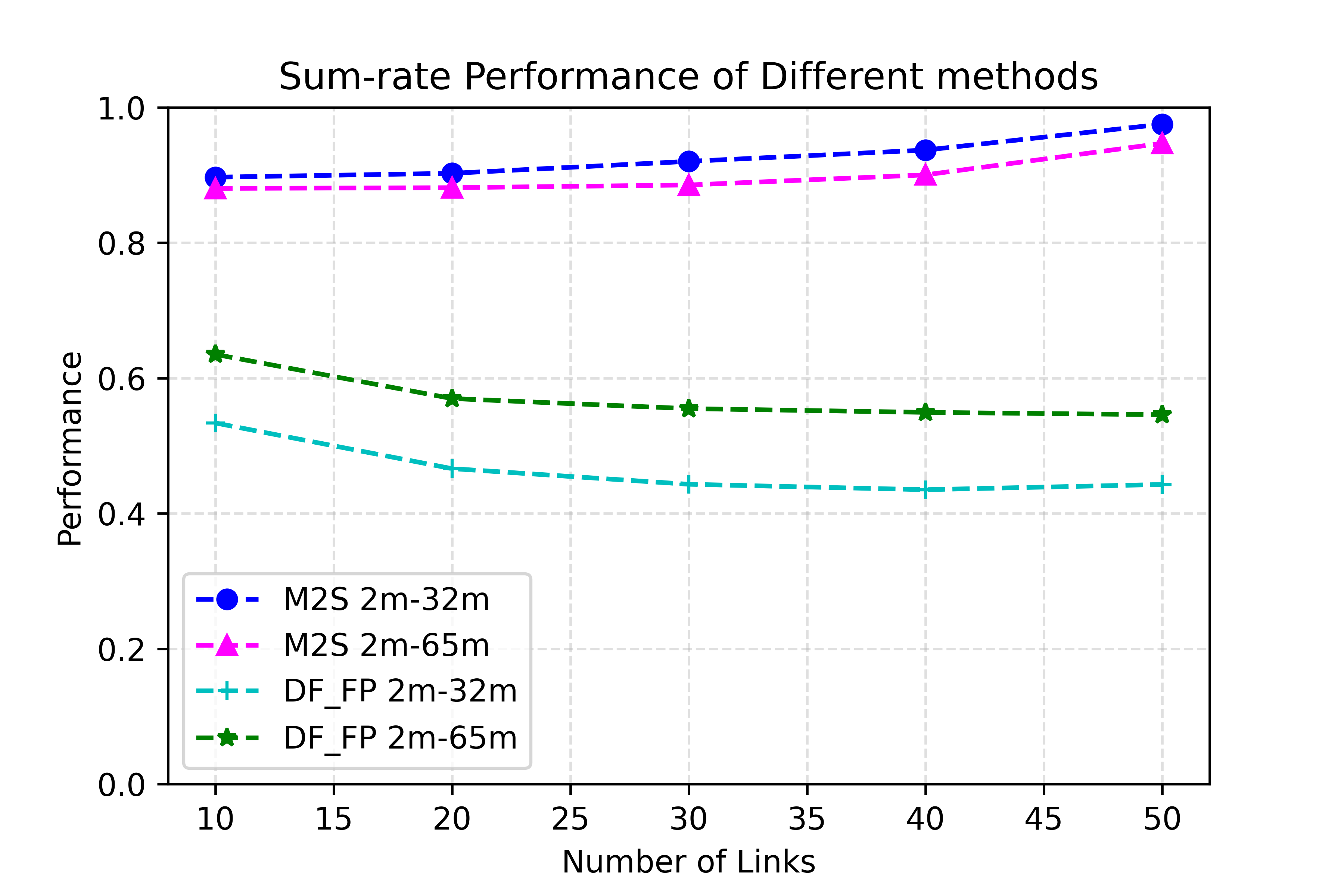}
    \caption{Sum-rate performance}
    \label{fig:sumrate}
\end{figure}

 As mentioned before, we choose  \textit{DPM} simulation for getting  accurate CSI and then use \textit{FPLinQ} algorithm for near-optimal scheduling as the $100\%$ performance baseline  (DPM-FP). In Fig. \ref{fig:sumrate}, ``M2S 2m-32m'' and ``M2S 2m-65m'' represent the sum-rate performance of \textit{Map2Schedule} under short ($d_i$ from 2m to 32m) and long ($d_i$ from 2m to 65m) link distance, respectively. The ``DF-FP 2m-32m'' and ``DF-FP 2m-65m'' represent the performance of the distance-based fading model (with only vehicle locations) and the \textit{FPLinQ} (DF-FP), which is the 100\% performance baseline of paper\cite{shen2017fplinq,lee2020graph}. The distance-based fading model applied the short-range outdoor model ITU-1411 with a distance dependent path-loss. \textit{Map2Schedule} can achieve over $90\%$ baseline performance on sum-rate metrics. However, the DF-FP method can only achieve approximately $50\%$. This verified that street layout together with vehicle locations in our model, can predict almost optimal scheduling policy while distance alone cannot provide satisfactory performance. 
\vspace{-0.2cm}
\vspace{-0.2cm}
\begin{table}[h]
\small
\centering  
\caption{Impacts of weighted BCE loss function}\label{result_table}
\begin{tabular}{|c|c|c|}
\hline
\diagbox{Metrics}{Method} & Map2Schedule& \makecell[c]{Original Graph \\Embedding  Model}  \\
\hline
Average accuracy & 79.08\% & \textbf{82.35\%}\\
\hline
Average sum-rate & \textbf{91.27\%} & 67.29\%\\
\hline
\makecell[c]{Sum-rate in \\50-long scenarios} & \textbf{94.71\%} & 53.83\%\\
\hline
\end{tabular}
\end{table}
\vspace{-0.2cm}
\subsection{Impacts of Weighted BCE Loss Function}
\par We compared the scheduling performance of the original graph embedding model and the second part of \textit{Map2Schedule} on the same CSI matrix input. As mentioned above, we deployed the weighted BCE loss function to address the scenario bias. We compare two metrics: average accuracy, which is the number of correctly predicted $x_i$ over $N$, and average sum-rate, which is the ratio of the objective function in Eq. (\ref{problem}) given predicted $x_i$ to DPM-FP. 
The result in Table \ref{result_table} reveals a slight reduction in average accuracy when using the weighted BCE loss function. This reduction can be attributed to the modification of the model, which makes the model focus more on ``good" link (high data rate) samples. Therefore, the accuracy on ``bad" link (low data rate) samples decreased, which contributed more to the total accuracy. However, the sum-rate metric significantly depends on the accuracy of ``good" links when ``good" links constitute a small proportion. As a result, \textit{Map2Schedule} with weighted BCE loss function was improved by an average of 35\% on the sum-rate metric and 76\% on the sum-rate metric in scenarios with 50 pair of longer links.

\subsection{Transferability between Different Scenarios}
\par
\begin{figure}[h]
    \centering
    \includegraphics[width=0.9\linewidth]{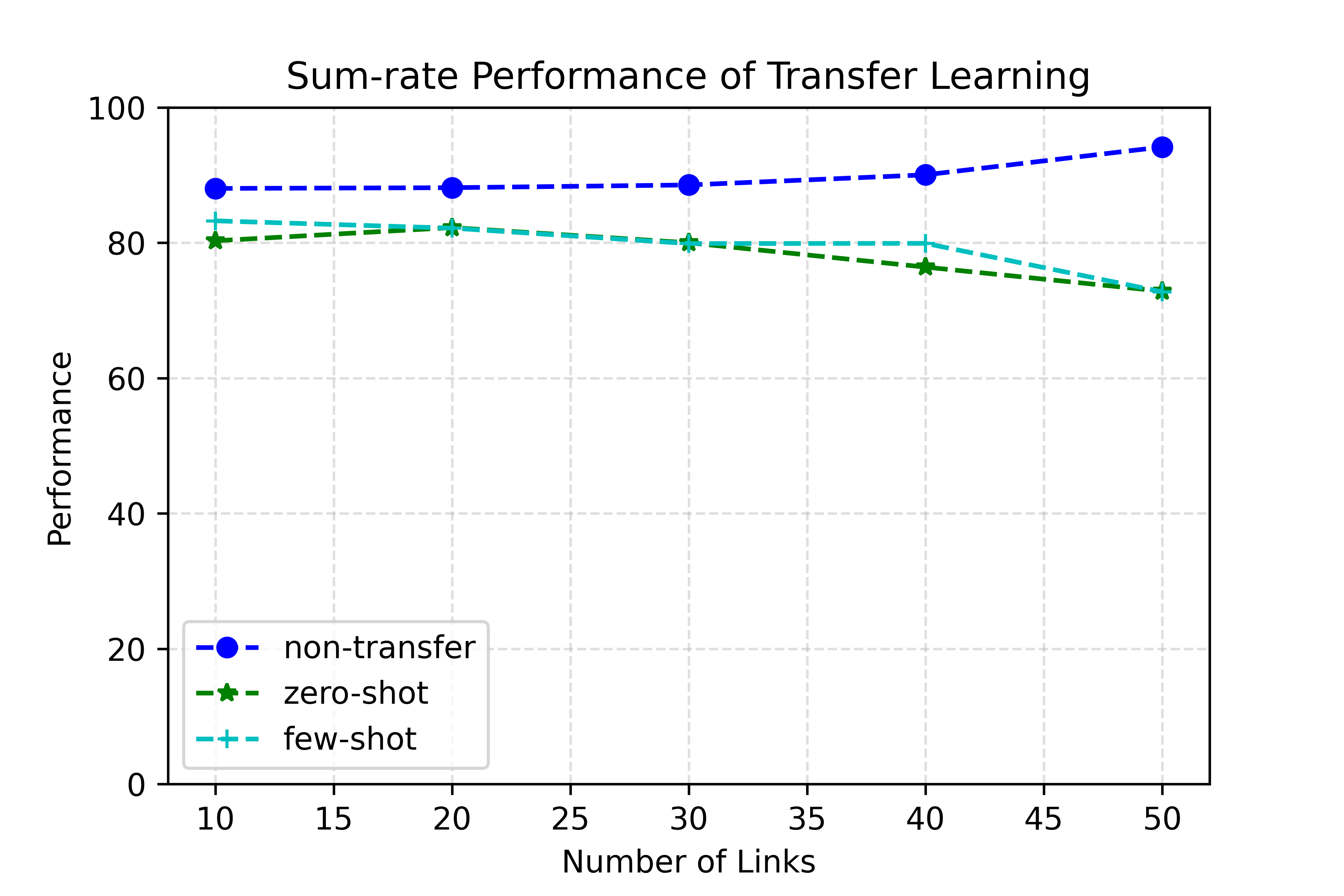}
    \caption{Sum-rate performance of transfer learning}
    \label{fig:transfer}
\end{figure}

 Transfer learning is a widely used ML technology that aims to improve the ML model performance on the target domain by transferring the knowledge from the original domain. In V2V scenarios, the vehicle density can change significantly within a single day, leading to very different CSI patterns. In this section, we tested our long-distance group model, which is the original domain knowledge, with the short-distance group data, which is the target domain. ``Zero-shot'' refers to without any view of the target domains, and ``few-shot'' refers to only viewing a few samples of the target domains. The ``non-transfer'' refers to the performance of models completely trained on the target domains. From the performance in Fig. \ref{fig:transfer}, we can find that our model exhibited promising transferability across different scenarios. However, the performance of transferred knowledge is lower than the original domain knowledge. The few shots slightly improved the performance, but further research on transfer learning is still needed for practical implementation.

\vspace{-0.2cm}
\vspace{-0.8cm}

\begin{figure}[h]
    \centering
    \includegraphics[width=0.9\linewidth]{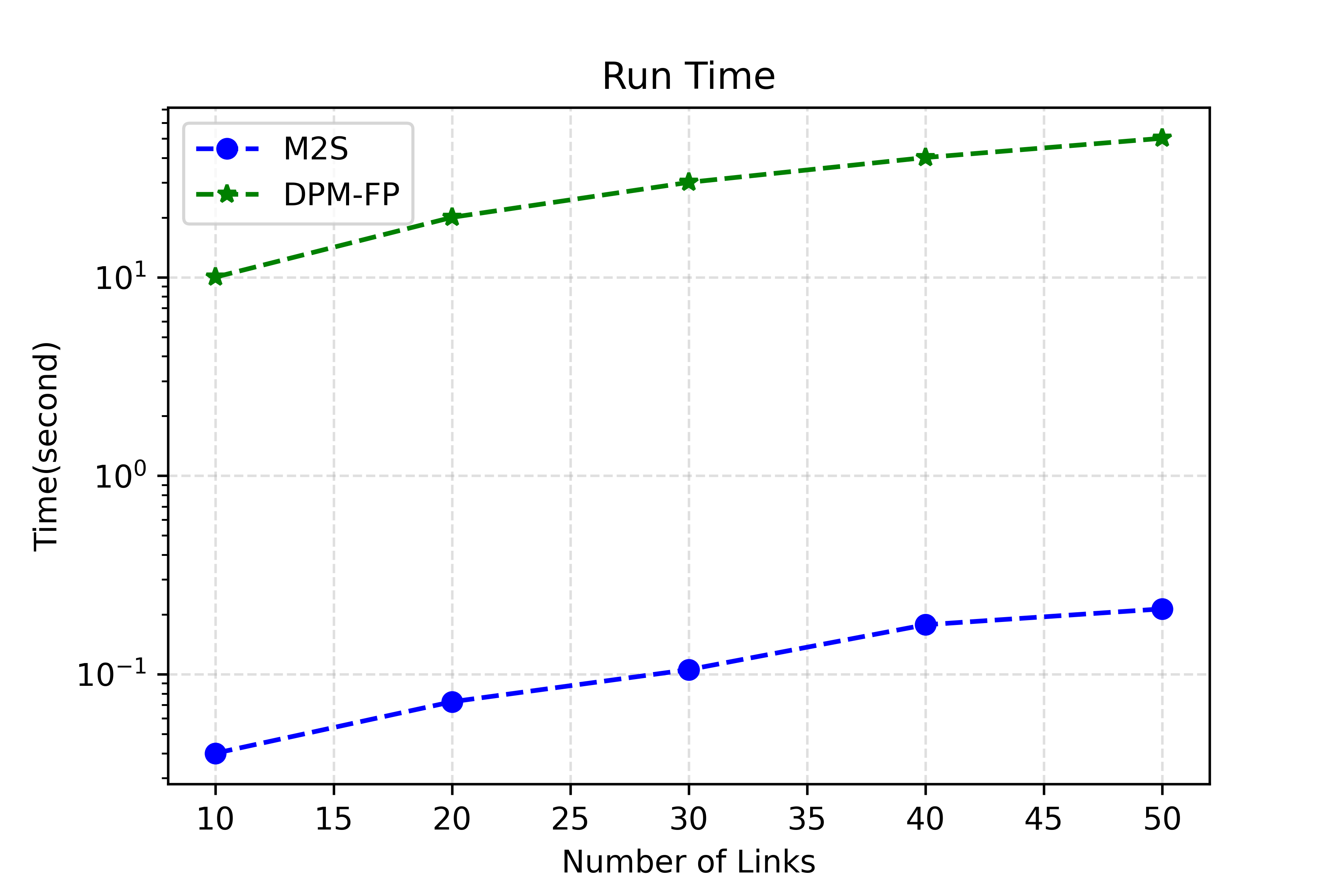}
    \caption{Run time comparison with baseline method}
    \label{fig:runtime}
\end{figure}

\vspace{-0.2cm}

\subsection{Run Time and the Computation Complexity}
All the experiments were conducted on our desktop with AMD Ryzen 5955WX processor and NVIDIA RTX A6000 graphic card. The average run time is shown in Fig. \ref{fig:runtime}. Specifically, the run time of DPM-FP is on the order of $10^{1}$ seconds, and the run time of \textit{Map2Schedule} is on the order of $10^{-1}$ second. Besides, further improvements can be achieved by carefully handling quantization in the data flow of embedding model. This result indicates \textit{Map2Schedule} has the desired low computational complexity and real-time feature.

\section{Conclusion}
In this paper, we proposed an ML-based end-to-end wireless link scheduling approach called \textit{Map2Schedule}, which specifically designed for urban V2V scenarios. \textit{Map2Schedule} can generate near-optimal link scheduling policy from vehicle locations and the city map. In urban environment, our approach remarkably outperforms those distance-based scheduling methods and competes with the state-of-the-art physical-model-based ones. Meanwhile, \textit{Map2Schedule} requires little computation resources, allowing it to meet the real-time requirement of V2V communications.

\bibliographystyle{IEEEtran}
\bibliography{lib}

\end{document}